\documentclass
[twocolumn,aps,prd,amsmath,showpacs,amssymb,nofootinbib]
{revtex4}

\usepackage[dvips]{graphicx}
\usepackage{bm}                      
\usepackage{mathptmx}                
\usepackage{dcolumn}                 

\def\bea{\begin{eqnarray}}
\def\eea{\end{eqnarray}}
\def\be{\begin{equation}}
\def\ee{\end{equation}}
\def\non{\nonumber}
\def\rra{\right\rangle}
\def\lla{\left\langle}
\def\eps{\varepsilon}
\def\sgm{\Sigma^-}

\def\la{\Lambda}

\def\ms{M_\odot}
\def\kv{\bm{k}}
\def\pv{\bm{p}}

\begin{document}

\title{Hybrid protoneutron stars with the Dyson-Schwinger quark model}


\author{H. Chen, M. Baldo, G. F. Burgio, and H.-J. Schulze}

\affiliation{
INFN Sezione di Catania, and Dipartimento di Fisica e Astronomia,
Universit\'a di Catania, Via Santa Sofia 64, 95123 Catania, Italy}


\begin{abstract}
We study the hadron-quark phase transition at finite temperature
in the interior of protoneutron stars,
combining the Dyson-Schwinger model for quark matter
with the Brueckner-Hartree-Fock approach for hadronic matter.
We discuss the dependence of the results on different nuclear three-body
forces and on details of the quark model.
A maximum mass exceeding two solar masses can be obtained with a strong
three-body force and suitable parameter values in the Dyson-Schwinger model.
With a hybrid configuration, the maximum mass of protoneutron stars is larger
then that of cold neutron stars,
such that a delayed collapse might be possible in principle.
\end{abstract}


\maketitle


\section{Introduction}

It is generally believed that a neutron star (NS) is formed as a result
of the gravitational collapse of a massive star ($M \gtrsim 8\ms$)
in a type-II supernova \cite{shapiro,bethe}.
Just after the core bounce, a protoneutron star (PNS)
is formed, a very hot and lepton-rich object,
where neutrinos are temporarily trapped.
The following evolution of the PNS is dominated by neutrino diffusion,
which results first in deleptonization and subsequently in cooling.
The star stabilizes at practically zero temperature, and no trapped
neutrinos are left \cite{burrows,rep}.

The dynamical transformation of a PNS into a NS
could be strongly influenced by a hadron-quark phase transition
in the central region of the star \cite{ponsevo,cooke,lugon,nemes}.
Calculations of PNS structure, based on a microscopic nucleonic
equation of state (EOS), indicate that for the heaviest PNS,
close to the maximum mass (about two solar masses),
the central baryon density reaches values larger than $1/\text{fm}^{3}$.
In this density range the nucleon cores (dimension $\approx 0.5\;\rm fm$)
start to touch each other, and it is likely that quark
degrees of freedom will play a role.

In previous articles \cite{proto,isen1,isen2}
we have studied static properties of PNS
using a finite-temperature hadronic EOS
including also hyperons \cite{pnsy}
derived within the Brueckner-Bethe-Goldstone (BBG) theory of nuclear matter.
An eventual hadron-quark phase transition was modeled within an extended
MIT bag model \cite{protomit,pnsymit}.
In the present work we consider a more sophisticated quark model,
the Dyson-Schwinger model (DSM)
\cite{Roberts:1994dr,Alkofer:2000wg,Roberts:2007jh,Roberts:2000aa,Chen:2011my}.

The study of hybrid stars is also important from another point of view:
Purely nucleonic EOS are able to accommodate fairly large (P)NS maximum masses
\cite{gle,bbb,akma,zhou,zhli},
but the appearance of hyperons in beta-stable matter could
strongly reduce this value \cite{zhli,mmy,carroll,djapo,nsesc}.
In this case the presence of non-baryonic, i.e., ``quark'' matter
would be a possible manner to stiffen the EOS and reach large NS masses.
Heavy NS thus would be hybrid quark stars.

The paper is organized as follows.
In section \ref{s:bhf} we review the determination of the baryonic
EOS in the BHF approach at finite temperature.
Section \ref{s:qm} concerns the QM EOS according to the DSM,
comparing also with the MIT bag model for reference.
In section \ref{s:res} we present the results regarding (P)NS structure,
combining the baryonic and QM EOS for beta-stable nuclear matter
via a Gibbs phase transition construction.
Section \ref{s:end} contains our conclusions.
\bigskip

\section{EOS of nuclear matter at finite temperature in the Brueckner approach}
\label{s:bhf}

The formalism proposed by Bloch \& De Dominicis \cite{bloch}
in order to solve the nuclear many-body problem at finite temperature
is the closest one to the Brueckner-Bethe-Goldstone (BBG) expansion \cite{balfer}.
In this approach the essential ingredient is the two-body scattering matrix $K$,
which along with the single-particle potential $U$
satisfies the self-consistent equations
\bea
 && \lla 1 2 | K(W) | 3 4 \rra =
 \lla 1 2 | V | 3 4 \rra
\label{eq:kkk}
\\&&\nonumber\hskip8mm
 + \text{Re} \sum_{5,6}
 \lla 1 2 | V | 5 6 \rra
 { [1-n(5)] [1-n(6)] \over
   W - E_{5} - E_{6} + i\eps }
 \lla 5 6 | K(W) | 3 4 \rra
\eea
and
\be
 U(1) = \sum_{2} n(2) \lla 1 2 | K(W) | 1 2 \rra_A \:,
\label{eq:ueq}
\ee
where $i=1,2,...$ generally denote momentum, spin, and hadron species.
Here $V$ is the two-body interaction,
$W = E_{1} + E_{2}$ represents the starting energy,
$E_{i} = k_i^2/2m_i + U_i(k_i)$ the single-particle energy, and
$n(i)$ is a Fermi distribution at finite temperature.
For given partial densities of all hadron species
$\rho_i$ $(i=n,p,\Lambda,\Sigma,\ldots)$
and temperature $T=1/\beta$,
Eqs.~(\ref{eq:kkk}) and (\ref{eq:ueq}) have to be solved
self-consistently along with the following equations for the auxiliary
chemical potentials $\tilde{\mu}_i$,
\be
 \rho_i = \sum_{k_i,s_i} n(i) =
 \sum_{k_i,s_i} {1\over e^{\beta (E_{i} - \tilde{\mu}_i)} + 1 } \:.
\label{e:ro}
\ee

The grand-canonical potential density $\omega$ in the Bloch-De Dominicis
framework can be written as the sum of a mean-field term
and a correlation contribution \cite{balfer},
\bea
 &&\hskip-4mm \omega = -\sum_i \bigg\{ \sum_k \left[ {1\over\beta}
 \ln\left( 1 +  e^{-\beta (E_i - \tilde{\mu}_i)}\right)
 + n_i(k) U_i(k) \right]
\\\nonumber&&\hskip0mm +
 {\frac{1}{2}} \int {dW\over 2\pi} e^{\beta(2\tilde{\mu}_i-W)}
 \,\text{Tr}_2\bigg(\!\arctan\big[{\cal K}(W) \pi\delta(H_0-W)
 \big]\bigg)\bigg\} \:,
\label{e:om}
\eea
where the trace $\rm Tr_2$ is taken in the space of
antisymmetrized two-body states and the two-body scattering matrix $\cal K$
is defined by
\be
  \lla 1 2 | {\cal{K}}(W) | 3 4 \rra  =
  \lla 1 2 | K(W) | 3 4 \rra
  \prod_{i=1,4}\! \sqrt{1-n(i)} \; .
\ee
By keeping only the first term in the expansion of the arctan function
\cite{balfer},
and neglecting a series of terms proportional to \hbox{$n(k)[1-n(k)]$}
(or powers of it),
which turn out to be negligible in the
temperature and density ranges relevant for (P)NSs,
then the correlation term simplifies and reduces to
\be
 \omega_c =
 {\frac{1}{2}} \sum_i\sum_{k} n_i(k) U_i(k) \:,
\label{e:om1}
\ee
which defines the grand-canonical potential in total analogy with the BBG
binding potential,
just using Fermi functions instead of the usual step functions
at zero temperature.

In this framework, the free energy density is then
\be
 f = \omega + \sum_i \rho_i \tilde{\mu}_i \:,
\label{e:f}
\ee
and all other thermodynamic quantities of interest can be computed from it,
namely the ``true" chemical potential $\mu_i$, pressure $p$,
entropy density $s$, and internal energy density $\eps$ read
\bea
 \mu_i &=& {{\partial f}\over{\partial \rho_i}} \:,
\\
 p &=& 
 \sum_i\mu_i \rho_i - f \:,
\\
 s &=& -{{\partial f}\over{\partial T}} \:,
\\
 \eps &=& f + Ts \:.
\eea

To solve Eq.~(\ref{eq:kkk}) one needs as input the interactions between hadrons.
In this paper we have used as nucleon-nucleon two-body force
the Argonne $V_{18}$ potential \cite{v18}.
As widely discussed in the literature,
in order to reproduce correctly the nuclear matter saturation point
$\rho_0 \approx 0.17~\mathrm{fm}^{-3}$, $E/A \approx -16$ MeV,
three-body forces (TBF) among nucleons are usually introduced.
In the BBG approach the TBF are reduced to density-dependent two-body forces
by averaging over the position of the third particle,
assuming that the probability of having two particles at a
given distance is reduced according to the two-body correlation function.
One should notice that both two- and three-body forces should be consistent,
i.e., use the same microscopic parameters in their construction.
In this paper, we will discuss results obtained following this approach
to TBF \cite{tbfmic},
which are based on the same meson-exchange parameters as the two-body potential.
For completeness, we also present results based on phenomenological models
for TBF, which are often adopted.
In particular, we use the phenomenological Urbana TBF \cite{uix},
here denoted by UTBF (UIX),
which consists of an attractive term due to two-pion exchange with excitation
of an intermediate $\Delta$ resonance,
and a repulsive phenomenological central term.

As far as the hyperonic sector is concerned,
we employ the Nijmegen soft-core nucleon-hyperon (NY) potentials NSC89,
which are fitted to the available experimental NY scattering data
\cite{nsc89,yamamoto}.
Those potentials have been widely used in the BBG zero-temperature calculations
\cite{zhli,mmy,sch98,vi00}.
It turns out that at zero temperature only $\la$ and $\sgm$ hyperons appear
in the NS matter up to very large densities.
We therefore also restrict the present study to these two hyperon species,
neglecting the appearance of thermal $\Sigma^0$ and $\Sigma^+$
and heavier hyperons.
We remind the reader that
hyperonic two- and three-body forces are neglected in our calculations,
due to the lack of experimental data.

In practice we employ empirical analytical parametrizations of the free energy
density, Eq.~(\ref{e:f}), as a function of partial densities and temperature,
as detailed in Refs.~\cite{isen2,pnsy,pnsymit}.

\section{Quark matter}
\label{s:qm}

\subsection{Dyson-Schwinger equations approach}

For the deconfined quark phase, we adopt a model based on the
Dyson-Schwinger equations of QCD,
which provides a continuum approach to QCD that can simultaneously address
both confinement and dynamical chiral symmetry breaking
\cite{Roberts:1994dr,Alkofer:2000wg}.
It has been applied with success to hadron physics in vacuum
\cite{Maris:2003vk,Roberts:2007jh,Eichmann:2008ef,Chang:2009zb},
and to QCD at nonzero chemical potential and temperature
\cite{Roberts:2000aa,Maas:2005hs,Fischer:2009wc,Chen:2008zr,Qin:2010nq}.
Recently efforts have been made to calculate the EOS for cold quark
matter and compact stars \cite{Zong:2008zzb,Klahn:2009mb,Chen:2011my},
and in this paper we extend the work of Ref.~\cite{Chen:2011my} to the
finite-temperature case.

Our starting point is QCD's gap equation for the quark propagator
$S(\pv,\omega_n;\mu,T)$
at finite quark chemical potential $\mu$ and temperature $T$,
which in the imaginary-time formalism reads
\footnote{
In our Euclidean metric:
$\{\gamma_\rho,\gamma_\sigma\} = 2\delta_{\rho\sigma}$;
$\gamma_\rho^\dagger = \gamma_\rho$;
$\gamma_5 = \gamma_4\gamma_1\gamma_2\gamma_3$;
$ab = \sum_{i=1}^4 a_i b_i$;
$\bm{a}\bm{b} = \sum_{i=1}^3 a_i b_i$;
and $P_\rho$ timelike $\Rightarrow$ $P^2<0$.}
\be
 S(\pv,\omega_n;\mu,T)^{-1} =
 i \bm{\gamma}\pv + i \gamma_4 (\omega_n+i\mu) + m_q
 + \Sigma(\pv,\omega_n;\mu,T)
\label{gendse}
\ee
with the self-energy expressed as
\bea
 \Sigma(\pv,\omega_n;\mu,T) &=&
 T\sum_{l=-\infty}^{\infty} \int\!\! \frac{d^3q}{(2\pi)^3} \,
 g^2 D_{\rho\sigma}(k;\mu,T)
\label{gensigma}
\\&& \times
 \frac{\lambda^a}{2}\gamma_\rho S(q;\mu,T) \Gamma^a_\sigma(q,p;\mu,T) \:,
\non
\eea
where $\omega_n=(2n+1)\pi T$ is the Matsubara frequency for quarks,
$p=(\pv,\omega_n)$,
$q=(\bm{q},\omega_l)$, and
${k=(\kv,\omega_{n-l}) = (\bm{p-q},\omega_n-\omega_l)}$.
Here, $g$ is the coupling strength,
$D_{\rho\sigma}(k;\mu,T)$ is the dressed gluon propagator,
and $\Gamma^a_\sigma(q,p;\mu,T)$ is the dressed quark-gluon vertex.
Moreover, $\lambda^a$ are the Gell-Mann matrices,
and $m_q$ is the current-quark mass.
As we employ an ultraviolet-finite model,
renormalization is unnecessary.

The kernel, Eq.~(\ref{gensigma}), depends on the gluon propagator and the
quark-gluon vertex.
Little is known about them at $T\neq 0$ and $\mu\neq 0$, hence we follow
the DSM that has been successfully applied to hadron physics at $\mu=T=0$.
For the quark-gluon vertex we use the ``rainbow approximation"
\be
  \Gamma_\sigma^a(q,p;\mu,T) = \frac{\lambda^a}{2}\gamma_\sigma \:,
\label{rainbowV}
\ee
which is the first term in a symmetry-preserving scheme to calculate
meson properties \cite{Bender:1996}.
For the gluon propagator in vacuum, we consider a Gaussian-type effective
interaction \cite{Alkofer:2002bp} in Landau gauge
\bea
g^2 D_{\rho \sigma}(k)
 = \frac{4\pi^2D}{\omega^6} \,
 k^2\, {\rm e}^{-k^2/\omega^2}\, P_{\rho\sigma}(k) \:,
\label{GluonAnsatz0}
\eea
with the transverse projector
$P_{\rho\sigma}(k) = \delta_{\rho \sigma}-{k_\rho k_\sigma\!/k^2}$.
It is a finite-width representation of the Munczek-Nemirovsky
model \cite{mn83} used in Ref.~\cite{Klahn:2009mb},
and expresses the long-range behavior of the
renormalization-group-improved effective interaction in
Refs.~\cite{Roberts:2007jh,Maris:1997tm,Maris:1999nt}.
Generally, at finite chemical potentials and temperatures,
the gluon propagator can be decomposed into electric and magnetic parts,
\bea
 g^2 D_{\mu\nu}(\kv,\Omega_n) =
 P_{\mu\nu}^T D_T(\kv^2,\Omega_n^2) +
 P_{\mu\nu}^L D_L(\kv^2,\Omega_n^2)
\eea
with the gluon Matsubara frequency $\Omega_n=2n\pi T$,
\be
 P_{\mu\nu}^T = \left\{ \begin{array}{ll}
 \delta_{\mu\nu} - {k_\mu k_\nu\!/\kv^2} & \mu,\nu=1,2,3\\
 0                                       & \mu,\nu=4
\end{array} \right. \:,
\label{ProjectorT}
\ee
and $P_{\mu\nu}^L = P_{\mu\nu} - P_{\mu\nu}^T$.
At $\mu=0$ and low temperatures,
there are indications \cite{Cucchieri:2007ta} that the gluon propagator is
insensitive to temperature.
In the interior of PNSs, the typical chemical potential is a few hundred MeV,
which is much larger than the typical temperature, a few tens MeV.
Therefore, we use the extended form \cite{Chen:2011my} of Eq.~(\ref{GluonAnsatz0})
\be
  D_T(\kv^2,\Omega_n) = D_L(\kv^2,\Omega_n) =
  \frac{4\pi^2D}{\omega^6} {\rm e}^{-\alpha\mu^2/\omega^2} k^2\,
  {\rm e}^{-k^2/\omega^2} \:,
\label{IRGsmu}
\ee
with $\alpha$ quantifying the asymptotic freedom at large chemical potential
but without any temperature effects.
As in Ref.~\cite{Chen:2011my},
we choose the set of parameters
$\omega=0.5\;{\rm GeV}$, $D=1\;{\rm GeV}^2$ \cite{Alkofer:2002bp}.
In the following, we consider three quark flavors $q=u,d,s$,
which are independent of each other in our model.
We take the current-quark masses $m_{u,d}=0$ for simplicity and
$m_s=115\;{\rm MeV}$  \cite{Alkofer:2002bp}.
The dependence of the $T=0$ EOS and the structure of NSs on $\alpha$
is discussed in Ref.~\cite{Chen:2011my}.
For $\alpha \approx 0$, the EOS of QM is too stiff to obtain
a hadron-quark phase transition in NSs,
while for $\alpha \approx \infty$,
the model is reduced to a free quark system at finite chemical potential.
Herein, we take two typical values $\alpha=2$ and $\alpha=4$ in our calculation.

\begin{figure}[t]
\centering
\includegraphics[scale=0.3, bb=0 390 770 1610,clip]{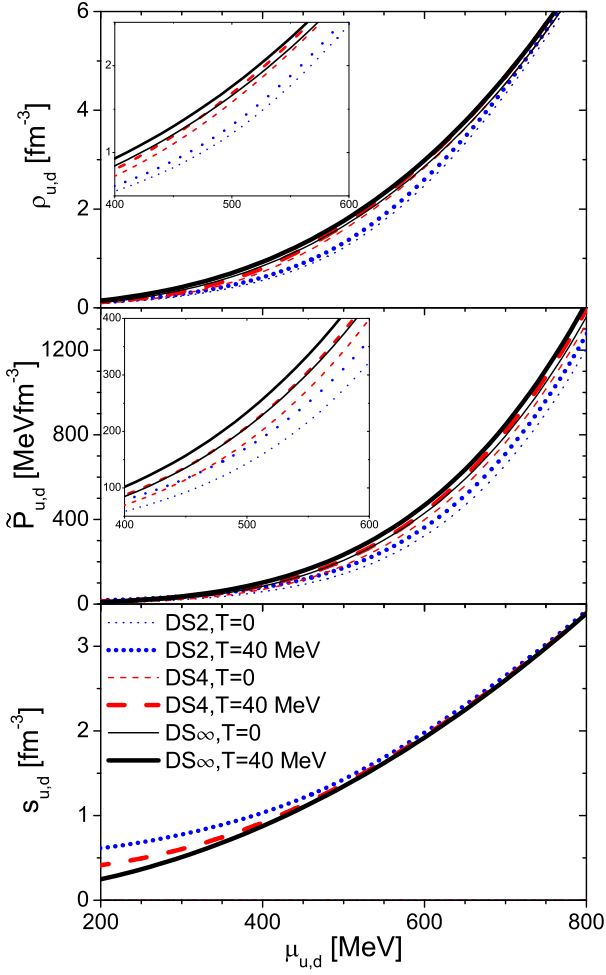}
\caption{
Quark number density (upper panel),
pressure $\tilde{P}_q$ (central panel),
and entropy density (lower panel)
for massless quarks at finite chemical potentials and temperatures.
Different curves correspond to different values of the parameter
$\alpha=2$ (dotted blue lines), 4 (dashed red lines),
$\infty$ (solid black lines) and
different temperatures $T=0$ (thin lines), $T=40$ MeV (thick lines).
All $s_{u,d}(\mu,T=0)=0$.
}
\label{fig:light}
\end{figure}

\begin{figure}[t]
\centering
\includegraphics[scale=0.3, bb=0 390 770 1610,clip] {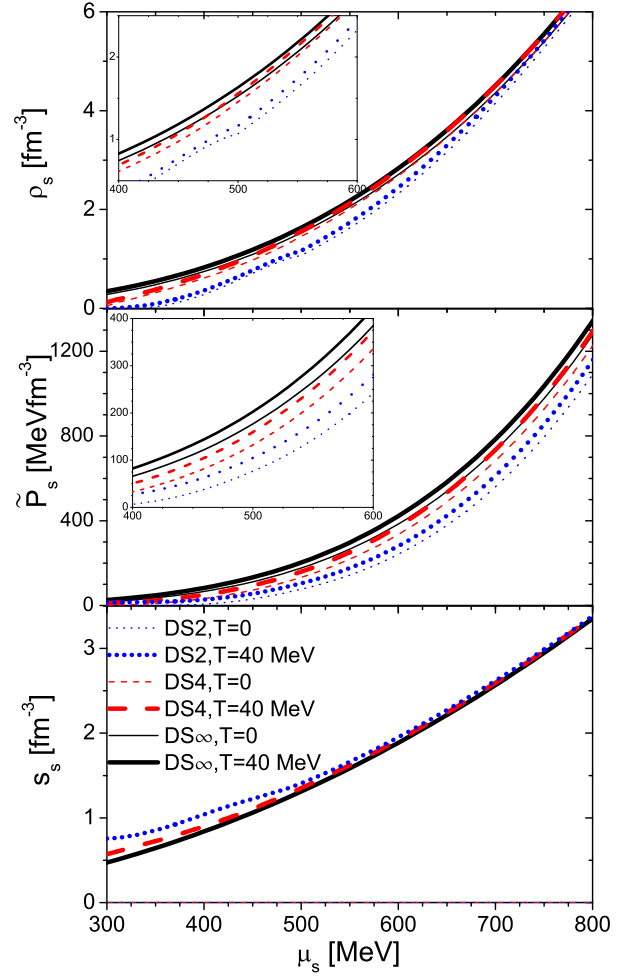}
\caption{
Same as Fig.~\ref{fig:light},
but for strange quarks with $m_s=115\;{\rm MeV}$.}
\label{fig:s}
\end{figure}

At finite chemical potential and temperature,
the quark propagator assumes a general form with rotational invariance
\bea
 S_q(\pv,\omega_n;\mu,T)^{-1} &=&
 i \bm{\gamma}\pv \;A(\pv^2,\omega_n;\mu_q,T)
 + B(\pv^2,\omega_n;\mu_q,T)
\non\\ &&
 +\; i \gamma_4(\omega_n+i\mu_q) \;C(\pv^2,\omega_n;\mu_q,T) \:.
\label{sinvp}
\eea
Here we ignore the possibilities of color superconductivity
\cite{Yuan:2006yf,Nickel:2006vf,Nickel:2006kc}
and other structures \cite{Roberts:2000aa}.
From the quark propagator, one can calculate the quark number density
and correspondingly obtain the pressure and EOS for QM
at zero temperature \cite{Chen:2008zr,Klahn:2009mb,Chen:2011my}.
In the case $T>0$, we express the single-flavor quark number density as
\bea
 \rho_q(\mu_q,T) &=& g \int\frac{d^3 p}{(2\pi)^3} \,
 \tilde{f}_q(|\pv|;\mu_q,T) \:,
\label{nqmu}
\\
 \tilde{f}_q(|\pv|;\mu_q,T) &=&
 \frac{T}{2}\sum_{n=-\infty}^{\infty} \,
 {\rm tr}_{\rm D}\left[-\gamma_4 S_q(\pv,\omega_n;\mu_q,T)\right] \:,
\label{nqmuf1}
\eea
where $g=2N_c=6$ is the quark degeneracy
and the trace is over spinor indices only.
Then, due to asymptotic freedom at large chemical potential $\mu_\text{UV}$,
the quark entropy density and the difference of pressure at $\mu_\text{UV}$
can be approximated as
\bea
 s_q(\mu_\text{UV},T) &\approx&
 s_q^\text{free}(\mu_\text{UV},T) \:,
\label{e:suv}
\\
\label{e:PqUV}
 P_q(\mu_\text{UV},T_1) - P_q(\mu_\text{UV},T_0) &\approx&
 \int_{T_0}^{T_1} dT s_q^\text{free}(\mu_\text{UV},T)\:,
\eea
where $s_q^\text{free}(\mu_{UV},T)$ is the quark entropy density
in a free quark system.
Therefore, if asymptotic freedom is approached continuously,
which is realized in the Wigner phase in our model with $\alpha>0.3$,
the single-flavor quark thermodynamic pressure
at finite chemical potential and temperature can be obtained as
\bea
 P_q(\mu_q,T) &=&
 P_q(\mu_{q,0},0) + \int_{\mu_{q,0}}^{\mu_{UV}}\! {\rm d}\mu' \,\rho_q(\mu',0)
\nonumber\\ &&\hspace{-10mm}
 + \int_0^T dT' s_q^\text{free}(\mu_\text{UV},T')
 + \int_{\mu_\text{UV}}^{\mu_q}\! {\rm d}\mu' \,\rho_q(\mu',T) \:.
\label{eq:pressure}
\eea
The total pressure for the quark phase is given by
summing contributions from all flavors.
For comparison with the bag model,
we write the pressure of the quark phase as
\be
 P_Q(\mu_u,\mu_d,\mu_s,T) \equiv
 \sum_{q=u,d,s}\tilde{P}_q(\mu_q,T) - B_\text{DS} \:,
\label{eq:PQ}
\ee
with the bag constant
\bea
 B_\text{DS} &\equiv& -\sum_{q=u,d,s} P_q(\mu_{q,0},0)\:.
 \label{e:Bdsm}
\eea
and the reduced pressure
$\tilde{P}_q(\mu_q,T)\equiv P_q(\mu_q,T) - P_q(\mu_{q,0},0)$
from the integrals of quark number density and entropy density.
As in Ref.~\cite{Chen:2011my}, we set $B_\text{DS} = 90$ MeV
with $\mu_{u,0}=\mu_{d,0}=0$ and $\mu_{s,0}$
as the value of the starting point of the
deconfined phase of strange quarks.
We choose $\mu_\text{UV}=1$ GeV in our calculation,
which is high enough as can be seen in the upper panels of
Figs.~\ref{fig:light} and \ref{fig:s}.

Correspondingly, the total baryon number density, entropy density,
internal energy density,
and free energy density for the quark phase are obtained as
\bea
 \rho_B &=&
 \frac{1}{3} \sum_{q=u,d,s}\rho_q(\mu_q,T)\:,
\\
 s_Q &=&
 \sum_{q=u,d,s}s_q(\mu_q,T) =
 \sum_{q=u,d,s}\frac{\partial \tilde{P}_q(\mu_q,T)}{\partial T}\:,
\\
 \epsilon_Q &=&
 -P_Q + \sum_{q=u,d,s}\mu_q \rho_q + T s_Q \:,
\\
 f_Q &=&
 \epsilon_Q - Ts_Q \:.
\label{eq:s}
\eea
In Figs.~\ref{fig:light} (for massless up and down quarks)
and \ref{fig:s} (for strange quarks)
we show the quark number density (upper panel),
pressure $\tilde{P}_q$ (central panel),
and entropy density (lower panel)
at finite chemical potentials and temperatures.
Both at zero and nonzero temperature,
when the chemical potential is about a few hundred MeV,
the quark number density in our model (DS2 and DS4) is lower than that of
the free quark system DS$\infty$.
This leads to a reduction of the pressure $\tilde{P}_q$.
Qualitatively, the increase of quark number density due to temperature effects
is similar in the DSM as in the free quark system.
The quark entropy density increases with the interaction strength parameter
$1/\alpha$ of the DSM.

In summary, both at zero and nonzero temperature,
the quark number density and pressure $\tilde{P}_q$ in the DSM are reduced
compared to the free quark system.
Since the typical temperatures are much smaller than the chemical potential,
the temperature effects are usually small.

\subsection{The MIT bag model at finite temperature}

We review briefly the description of the bulk properties of
uniform QM at finite temperature
within the MIT bag model \cite{chodos}.
In its simplest form,
the quarks are considered to be free inside a bag and the thermodynamic
properties are derived from the Fermi distribution, i.e.,
\bea
 \tilde{f}_q(k;\tilde{\mu}_q,T) &=& f^+_q(k) - f^-_q(k) \:,
\label{nqmuf1bag} \\
 f^\pm_q(k) &=& \frac{1}{ 1+\exp[(E_q(k) \mp \tilde\mu_q)/T] }\:,
\label{distf}
\eea
with $E_q(k) = \sqrt{m_q^2+k^2}$
and $\tilde{\mu}_q$ being the auxiliary quark chemical potentials.
The quark number density is obtained from Eq.~(\ref{nqmu}) and
the energy density and free energy density for the quark phase are given as
\bea
 \eps_Q &=& B + \sum_q g\int\!\!\frac{d^3k}{(2\pi)^3}
 \left[f^+_q(k) + f^-_q(k)\right] E_q(k) \:,
\label{e:epsq}
\\
 f_Q &=& \eps_Q - T\sum_q s^\text{free}_q \:,  
\label{e:fq}
\eea
where
$B$ is the bag constant and
$s^\text{free}_q$ the entropy density of a noninteracting quark gas.
The `true' quark chemical potential, entropy density, and pressure
are given by the general relations
\bea
 \mu_q &=&
 \frac{\partial f_Q(\rho_u,\rho_d,\rho_s,T)}{\partial \rho_q} \:,
\label{muq:mitb}
\\
 s_Q &=& \frac{\partial f_Q(\rho_u,\rho_d,\rho_s,T)}{\partial T} \:,
\label{sq:mitb}
\\
 p_Q &=& \sum_q \mu_q \rho_q - f_Q \:.
\label{Pq:mitb}
\eea
The corresponding expressions at $T=0$ can be obtained
by eliminating the antiparticles and
substituting the particle distribution functions by the usual step functions.
We have used massless $u$ and $d$ quarks, and $m_s=150$ MeV.

It has been found \cite{nsquark} that within the MIT bag model
(without color superconductivity) with a constant bag parameter
$B \simeq \rm 90~MeV~fm^{-3}$,
the maximum mass of a NS cannot exceed a value of about 1.6 solar masses.
Indeed, the maximum mass increases as the value of $B$ decreases,
but too small values of $B$ are incompatible
with a hadron-quark transition at baryon density $\rho_B >$ 2--3 $\rho_0$ in
nearly symmetric nuclear matter,
as demanded by heavy-ion collision phenomenology.
(These baryon densities are usually reached in numerical simulations \cite{hic}
of heavy ion collisions at intermediate energies
without yielding indications of `exotic' physics.)

In order to overcome these problems, we have introduced
in a phenomenological manner
a density-dependent bag parameter $B(\rho_B)$ \cite{nsquark}.
This allows one to lower the value of $B$ at large density,
providing a stiffer QM EOS and increasing the value of the NS maximum mass,
while at the same time still fulfilling the condition of
no phase transition below $\rho_B \approx 3 \rho_0$ in symmetric matter.
In the following we present results based on the MIT model using both
a constant value of the bag parameter,
and a Gaussian parametrization for the density dependence,
\be
 {B(\rho_B)} = B_\infty + (B_0 - B_\infty)
 \exp\left[-\beta\Big(\frac{\rho_B}{\rho_0}\Big)^2 \right]
\label{eq:param} \ee
with $B_\infty = 50\;\rm MeV~ fm^{-3}$, $B_0 =
400\;\rm MeV~ fm^{-3}$, and $\beta=0.17$,
see Ref.~\cite{nsquark}.
In this paper, we disregard possible dependencies of the bag parameter
on temperature and individual quark densities.
For a more extensive discussion of this topic, the reader is referred to
Refs.~\cite{nsquark,cdm}.

\section{Results and discussion}
\label{s:res}

\subsection{Matter in beta equilibrium}

\begin{figure*}[t]
\vspace{-15mm}
\includegraphics[scale=0.65,clip]{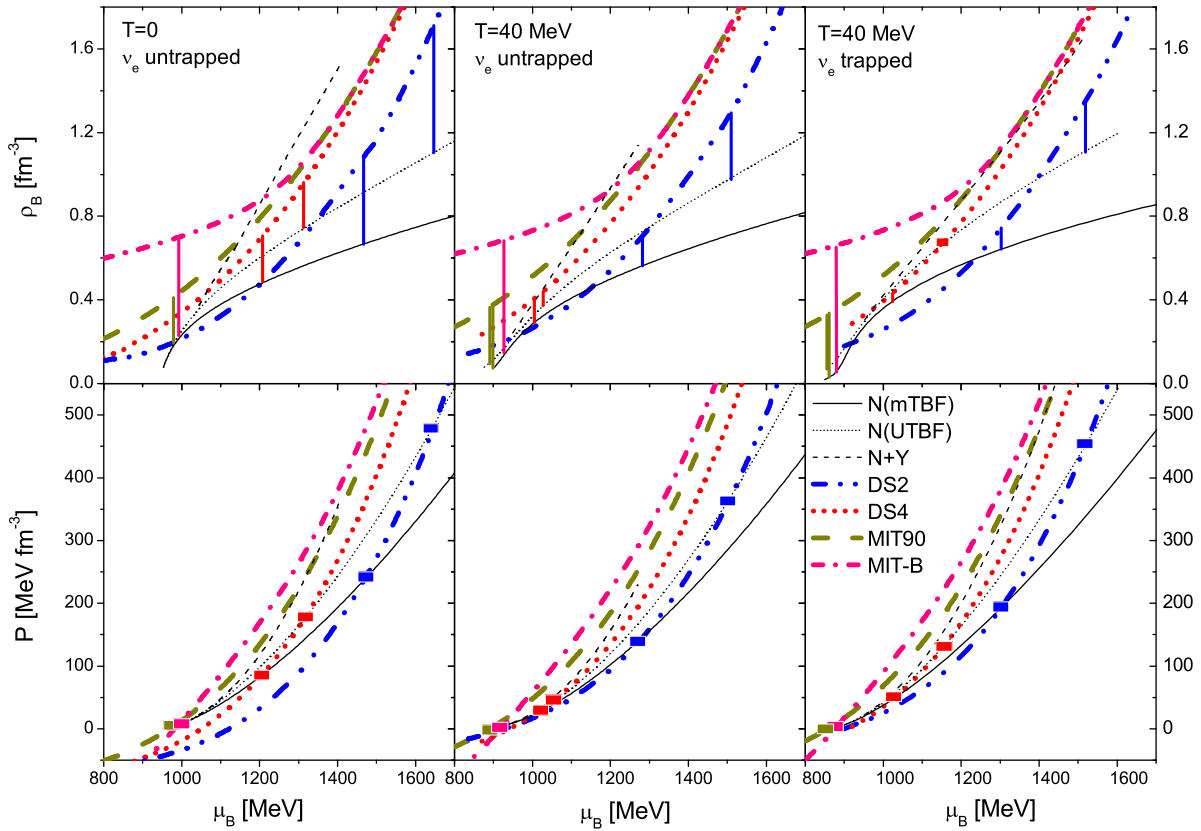}
\vspace{-25mm}
\caption{(Color online)
Baryon density (upper panels) and pressure (lower panels) as
function of the baryon chemical potential
for cold untrapped matter (left panels),
untrapped matter at $T=40$ MeV (central panels), and
trapped matter at $T=40$ MeV (right panels).
See text for details.
}
\label{f:max}
\end{figure*}

In neutrino-trapped beta-stable nuclear matter
the chemical potential of any particle $i=n,p,\la,\sgm,u,d,s,l$
is uniquely determined by the conserved quantities
baryon number $B_i$, electric charge $Q_i$,
and weak charges (lepton numbers) $L^{(e)}_i$, $L^{(\mu)}_i$
with corresponding chemical potentials $\mu_B,\mu_Q, \mu_{L_e}, \mu_{L_\mu}$:
\be
 \mu_i = B_i\mu_B + Q_i\mu_Q
 + L^{(e)}_i \mu_{L_e} + L^{(\mu)}_i \mu_{L_\mu}  \:,
\label{e:mufre}
\ee
where $B_i,Q_i,L^{(e)}_i,L^{(\mu)}_i$
are the corresponding charges of each particle.
The relations of chemical potentials and densities for hadrons and quarks
are given in the previous sections, and we treat leptons as free fermions.
With such relations, the whole system in each phase can be solved
for a given baryon density $\rho_B=\sum_i B_i \rho_i $,
imposing the charge neutrality condition
\be
 \sum_i Q_i \rho_i = 0
\label{e:neutral}
\ee
and lepton number conservation
\be
 Y_l = \sum_i L_i^{(l)} \frac{\rho_i}{\rho_B} \:.
\label{e:lepfrac}
\ee
We fix the lepton fractions to $Y_e=0.4$
for neutrino-trapped matter at $T=40\;\text{MeV}$,
and we neglect muons and muon-neutrinos due to their low fractions,
hence $Y_{\mu}=0$ .
When the neutrinos $\nu_e$ are untrapped,
the lepton number is not conserved any more,
the density and the chemical potential of $\nu_e$ vanish,
and the above equations simplify accordingly.


In Fig.~\ref{f:max} we display
the relations between baryon chemical potential, baryon density, and pressure
for the pure hadron and quark phases of beta-stable matter.
Thin curves represent the hadronic phases obtained
with the Argonne $V_{18}$ NN potential
supplemented by microscopic TBF (solid black)
or phenomenological Urbana TBF (dotted black),
and including also hyperons (dashed black).
The thick curves are the results for beta-stable QM, i.e.,
the Dyson-Schwinger results with $\alpha=2$ (dash-dot-dotted blue)
and $\alpha=4$ (dotted red), whereas
the other two lines represent the results obtained with the MIT bag model,
using either a constant bag parameter $B=\rm 90~MeV~fm^{-3}$ (dashed green)
or a density-dependent B($\rho$) (dash-dotted pink).
The left-hand panels display results for cold beta-stable matter,
the central panels for hot ($T=40$ MeV) and untrapped matter,
and the right-hand panels for hot and trapped matter.
The crossing points
of the baryon and quark pressure curves represent the transition between
baryon and QM phases under the Maxwell construction.
The projections of these points (vertical lines) on the baryon and quark density
curves in the upper panels indicate the corresponding transition densities
from low-density baryonic matter to high-density QM.

We notice that the phase transition from hadronic to QM occurs at low values
of the baryon chemical potential when the MIT bag model is used to describe
the quark phase.
This holds true irrespective of the EOS adopted for the hadronic matter,
since the TBF effects are only important at high densities.
On the contrary, with the DSM for QM
the phase transition occurs at higher values of the density,
since the EOS is generally stiffer than the hadronic one.
In this case, the onset and the width of the density range characterizing
the phase transition depend strongly on the EOS used for the hadronic phase.

Thermal effects play an important role,
since they shift the onset of the phase transition to lower values
of the baryon density.
Neutrino trapping lowers even more the transition density in the MIT case,
but increases it for the DSM.
This is because of the different behavior of the nucleonic EOS
at low and high densities:
At low density, corresponding to the transition density in the MIT case,
neutrino trapping increases the pressure and decreases the transition density.
At large density, corresponding to the transition density in the DSM case,
nuclear matter has a larger symmetry energy,
and the decrease of nucleon pressure overwhelms the increase from the leptons.
Therefore, neutrino trapping decreases the total pressure
and increases the transition density.
We also notice that with the DSM no phase transition exists
if the hadronic phase contains hyperons and the EOS is very soft,
or if the parameter $\alpha$ is too small and the EOS of QM is very stiff,
in analogy with the zero-temperature case \cite{Chen:2011my}.

\subsection{The hadron-quark phase transition}

\begin{figure*}[t]
\vspace{-15mm}
\includegraphics[scale=0.65,clip]{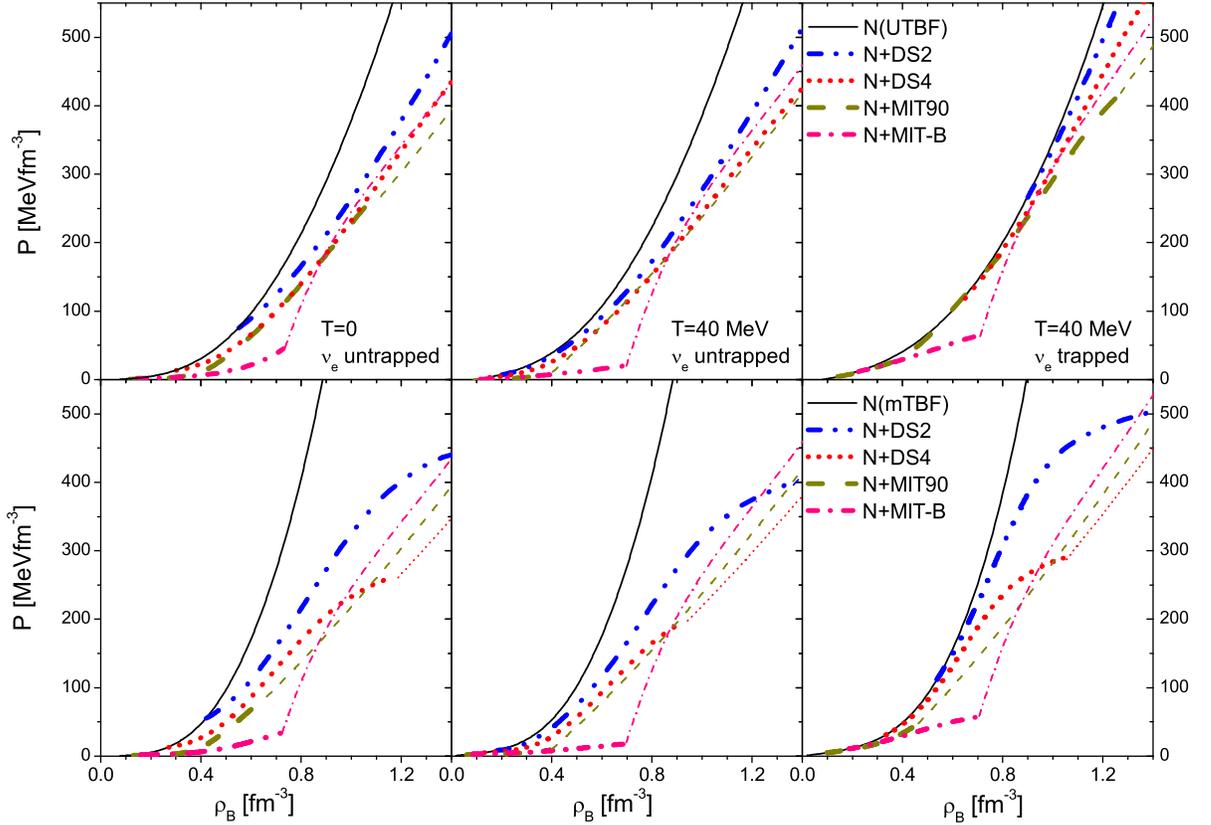}
\vspace{-25mm}
\caption{(Color online)
Pressure vs.~baryon density of NS matter
with the Gibbs phase transition construction for different models.
Phenomenological (upper panels) or microscopic (lower panels) TBF
are used for the hadronic EOS.
The pure hadron or quark phases are marked with thin curves
and the mixed phase region with thick curves.
The pure quark phase regions with the DSM in the upper panel
and with DS2 in the lower panel lie beyond the range we plot.
}
\label{f:gle_uix}
\end{figure*} 

We found in Ref.~\cite{Chen:2011my} that hybrid NS with the DSM
obtained using the Maxwell construction are unstable.
We therefore study in the following the phase transition at finite temperature
with the Gibbs construction \cite{gle,maru,glen},
which determines a range of baryon densities where both phases coexist,
yielding an EOS containing a pure hadronic phase, a mixed phase,
and a pure QM region.
In the Gibbs construction, as suggested by Glendenning \cite{glen},
the stellar matter is treated as a two-component system,
and therefore is parametrized by two chemical potentials.
Usually the pair ($\mu_e, \mu_n$), i.e.,
electron and baryon chemical potential, is chosen.
By imposing mechanical, chemical, and thermal equilibrium,
it turns out that the pressure is a monotonically increasing function
of the density,
at variance with the Maxwell construction,
where a plateau in the pressure vs.~density plane exists,
thus characterizing the phase transition.
We note that the Gibbs treatment is the zero surface tension limit
of the calculations including finite-size effects \cite{maru,yasu}.
The Gibbs phase transition has been widely studied in the literature,
and the formalism will not be repeated here.

In Fig.~\ref{f:gle_uix} we display results for the EOS involving the
Gibbs hadron-quark phase transition,
comparing calculations using phenomenological (upper panel)
and microscopic (lower panel) TBF.
We adopt similar conventions as in Fig.~\ref{f:max} for the QM EOS,
but with thin lines indicating the pure hadron or quark phases
and thick lines the mixed phase region.
Similar as in Fig.~\ref{f:max} for the phase transition
with the Maxwell construction,
thermal effects alone (central panels) shift the onset of the mixed phase
to smaller density,
while neutrino trapping delays the onset.
Taking all effects into account,
the EOS comprising the phase transition turns out to be softer
than the hadronic one.

Comparing the phase transition constructed with the DSM
and that obtained using the MIT bag model,
they turn out to be quite different,
as predicted from Fig.~\ref{f:max}.
In the former case, the onset of the phase transition is shifted to a
larger baryonic density and the mixed phase is extended to a much larger region.
Different nuclear EOSs also affect the onset and the width of the mixed phase,
in particular when the DS model is used.
In the case of a stiff EOS for the hadronic phase with the microscopic TBF,
the onset of the mixed phase is at lower density
and the mixed phase has a smaller region,
compared to the case of phenomenological TBF.
These all have consequences for the internal structure of a PNS,
as discussed in the next subsection.

\subsection{Proto-neutron star structure}

\begin{figure*}[t]
\vspace{-15mm}
\includegraphics[scale=0.65,clip]{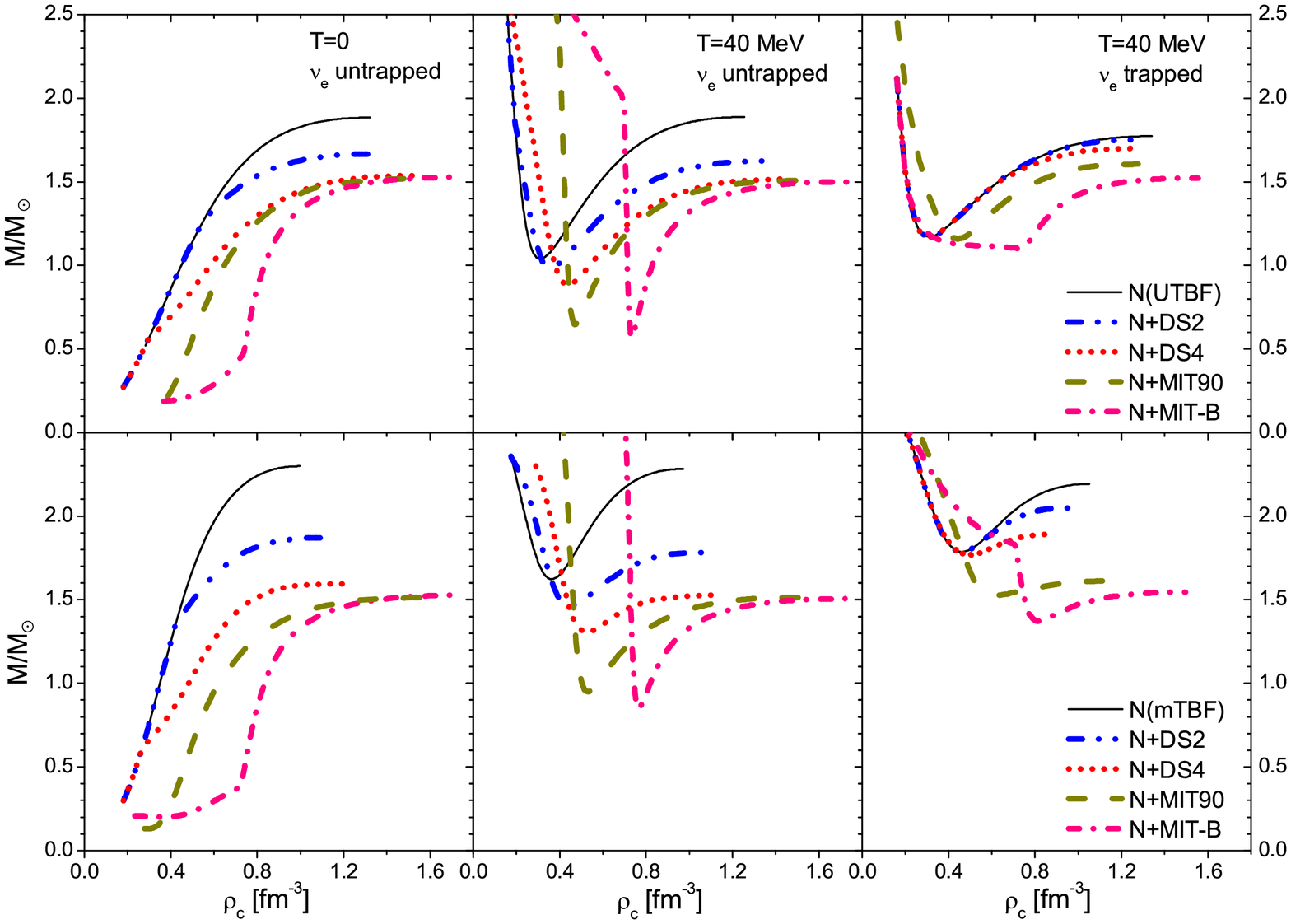}
\vspace{-25mm}
\caption{(Color online)
NS gravitational mass vs.~central baryon density.
Phenomenological (upper panels) or microscopic (lower panels) TBF
are used for the hadronic EOS.
}
\label{f:mrho1}
\end{figure*}

The stable configurations of a (P)NS can be obtained from the
well-known hydrostatic equilibrium equations
of Tolman, Oppenheimer, and Volkov \cite{shapiro}
for pressure $p(r)$ and enclosed mass $m(r)$,
\bea
 {dp\over dr} &=& -\frac{Gm\eps}{r^2}
 \frac{ \big( 1 + p/\eps \big) \big( 1 + 4\pi r^3p/m \big)}
 {1-2Gm/r} \:,
\label{tov1:eps}
\\
 \frac{dm}{dr} &=& 4\pi r^{2}\eps \:,
\label{tov2:eps}
\eea
once the EOS $p(\eps)$ is specified ($G$ is the gravitational constant).
For a given central value of the energy density, the numerical integration of
Eqs.~(\ref{tov1:eps}) and (\ref{tov2:eps})
provides the mass-central density relation.
In the low-density range, where nucleonic clustering sets in,
we cannot use the BHF approach,
and therefore we join \cite{isen2} the BHF EOS to the
finite-temperature Shen EOS \cite{shen},
which is more appropriate at densities below $\rho \leq 0.07\;\text{fm}^{-3}$,
since it does include the treatment of finite nuclei.

Our results, using the different EOSs introduced in the previous section,
are displayed in Fig.~\ref{f:mrho1},
which shows the gravitational mass
(in units of the solar mass $M_\odot=1.98\times 10^{33}$g)
as a function of the central baryon density.
We use the same conventions as in the previous figures:
The black solid curves represent the calculations performed for
purely hadronic matter,
using either phenomenological (upper panels) or microscopic (lower panels) TBF.
The colored broken curves denote stellar configurations of hybrid (P)NSs
with neutrino-free (left and central panels) and neutrino-trapped matter
(right panels).

We notice that the value of the maximum mass decreases in neutrino-free
matter due to thermal effects,
both in the purely hadronic case
and including the hadron-quark phase transition,
where the decrease depends on the EOS used for the quark phase
and turns out to be more relevant for the DS model.
On the contrary,
neutrino trapping
decreases further the maximum mass of purely hadronic stars,
but increases the value of the maximum mass of hybrid stars,
overcoming the thermal effect.
Therefore in this case a delayed collapse scenario is possible,
just as for hyperon stars \cite{pnsy}.

One also notices a dependence on the EOS used for the hadronic phase,
which is more important for the DS model,
where the QM onset takes place in a range of densities where
TBF play an important stiffening role,
and this explains the different values of the maximum mass.
On the contrary,
with the MIT bag model the transition takes place at low baryon densities where
the different TBF behave similarly, hence the small influence.
The proper values of the maximum mass are summarized in Table~\ref{t:mass}
and a maximum mass exceeding 2 $M_\odot$ for the hybrid configurations
can be obtained with the microscopic TBF
and suitable values of the parameter $\alpha$ in the DSM.

\begin{table}[t]
\setlength{\tabcolsep}{5pt}
\caption{
Properties of (P)NS maximum mass configurations.
Microscopic TBF are used.
}
\begin{ruledtabular}
\begin{tabular}
{l|l|ddd}
\multicolumn{2}{l|}{}
& \multicolumn{1}{c}{$M/M_\odot$}
& \multicolumn{1}{c}{$R$ (km)}
& \multicolumn{1}{c}{$\rho_c/\rho_0$} \\
\hline
        & N (mTBF)  & 2.30 & 10.85 & 5.86 \\
 untrapped  & DS2  & 1.87 & 11.02 & 6.54 \\
 T=0        & DS4  & 1.60 & 10.61 & 7.09 \\
      & MIT, B=90  & 1.51 & 9.02 & 9.27 \\
& MIT, B=B($\rho$) & 1.53 & 8.37 & 10.23 \\
\hline
        & N (mTBF)   & 2.28 & 13.14 & 5.72 \\
 untrapped  & DS2   & 1.78 & 14.68 & 6.47 \\
 T=40 MeV   & DS4   & 1.53 & 15.46 & 6.65 \\
 	   & MIT, B=90  & 1.51 & 11.62 & 8.87 \\
 & MIT, B=B($\rho$) & 1.51 & 10.37 & 10.13 \\
\hline
        & N (mTBF)   & 2.19 & 12.85 & 6.18 \\
 trapped    & DS2   & 2.05 & 15.33 & 5.70 \\
 T=40 MeV   & DS4   & 1.89 & 17.55 & 5.26 \\
 	   & MIT, B=90  & 1.61 & 15.82 & 6.98 \\
 & MIT, B=B($\rho$) & 1.55 & 13.37 & 8.76 \\
\end{tabular}
\end{ruledtabular}
\label{t:mass}
\end{table}

\section{Conclusions}
\label{s:end}

Extending the work of Ref.~\cite{Chen:2011my} to finite temperature,
we have studied the properties of hybrid PNSs
based on a hadronic BHF EOS involving different nuclear TBF
joined via a Gibbs construction to the DSM quark EOS.
This is only possible if hyperons are excluded from the hadronic phase.

We find a sizeable dependence of the results,
in particular for the PNS maximum mass,
on the nuclear TBF and on the quark model employed.
The DSM model allows to reach larger masses than the MIT model,
depending on the value of the interaction strength parameter.

In all cases finite temperature reduces slightly the maximum mass,
but neutrino trapping increases it sufficiently for hybrid stars,
such that a delayed collapse might be possible in principle.

\section*{Acknowledgments}

This work was partially supported by CompStar,
a Research Networking Programme of the European Science Foundation,
and by the MIUR-PRIN Project No. 2008KRBZTR.


\end{document}